\documentclass[a4paper,12pt]{article}
\usepackage{CJK}
\usepackage{color}
\usepackage{graphicx}
\usepackage{epstopdf}
\usepackage{amssymb}
\usepackage{amsfonts}
\usepackage{amsmath}
\usepackage{mathrsfs}
\usepackage[numbers,sort&compress]{natbib}
\usepackage{epsf}
\usepackage{abstract}
\usepackage{authblk}
\usepackage{bm}
\long\def\begincomment#1\endcomment{}

\usepackage{xcolor} 

\begin{document}
\setlength{\baselineskip}{7.5mm}

\title { \textbf{ A node-based SIRS epidemic model with infective media on complex networks }
}

\author[ a]{Leyi Zheng}
\author[, a, b]{Longkun Tang \footnote{Corresponding Author: tomlk@hqu.edu.cn }}
\affil[a]{\small \emph{Fujian Province University Key Laboratory of Computation Science}, \authorcr 
          \emph{School of Mathematical Sciences, Huaqiao University, Quanzhou {\rm 362021}, China}.}
\affil[b]{\small \emph{Department of Mathematics {\rm \&} Statistics, Georgia State University, Atlanta {\rm 30303}, USA}.}

\date{}
\maketitle{}

\begin{abstract}
\noindent In this paper, we focus on the node-based epidemic modeling for networks, introduce the propagation medium and propose a node-based Susceptible-Infected-Recovered-Susceptible (SIRS) epidemic model with infective media. Theoretical investigations show that the endemic  equilibrium is globally asymptotically stable. Numerical examples of three typical network structures also verify the theoretical results. Furthermore, Comparison between network node degree and its infected percents implies that there is a strong positive correlation between both,
namely, the node with bigger degree is infected with more percents. Finally, we discuss the impact of the epidemic spreading rate of media as well as the effective recovered rate on the network average infected state. Theoretical and numerical results show that
(1) network average infected percents go up (down) with the increase of the infected rate of media (the effective recovered rate); (2) the infected rate of media has almost no influence on network average infected percents for the fully-connected network and NW small-world network; (3) network average infected percents decrease exponentially with the increase of the effective recovered rate, implying that the percents can be controlled at low level by an appropriate large effective recovered rate.

\vskip 0.25cm \noindent
\noindent{}{\textbf{Keywords:}} SIRS; node-based;\ propagation medium;\ complex network;\ epidemic spreading;\ stability.
\end{abstract}

\section{Introduction}
With the development of network science, the mathematical modeling of epidemic spreading has involved in a research area across many disciplines including mathematical biology, physics, social science, computer and information science, and so on. On the basis of classical epidemic spreading models, such as, Susceptible-Infected-Susceptible (SIS) model, Susceptible-Infected-Recovered (SIR) model, and Susceptible-Infected-Recovered-Susceptible (SIRS) model,  a variety of epidemic spreading models \cite{Pastor2001, Mishra2007, Yuan2008, Newman2002, Selley2015, Liu2017, Wei2015, WeiWu2018, Pastor2015, Mieghem2009, YoussefM2011, Plos one2015, LX Yang2017} in networks were developed. Investigations on these models have important significance in public-health domain, especially in infectious disease epidemiology, by providing a number of interesting and unexpected behaviors.

The theoretical studies of epidemic spreading models in complex networks rely mostly on the mean-field theory approaches, especially on
degree-based mean-field (DBMF) theory which was the first theoretical approach presented for the analysis of general dynamical processes on complex networks \cite{Pastor2015}. This approach assumes that all nodes of degree $k$ are statistically equivalent, and any given vertex of degree $k$ is connected with the same probability to any node of degree $k\,'$. Therefore, the epidemic spreading model based on DBMF theory depends in general on the statistical topological properties of the underlying networks instead of the whole network structure, resulting into the loss of detailed features of network topologies such that it is difficult to deeply understand the effect of network structures on the disease (or information) propagation. To the best of our knowledge, in 2009, Mieghem et al. \cite{Mieghem2009} firstly proposed the continuous-time node-based SIS epidemic spreading model for understanding the influence of network characteristics on epidemic spreading. Youssef and Scoglio \cite{YoussefM2011} established a new individual-based SIR model with the whole description of network structures. Very recently, Yang et al. \cite{Plos one2015} suggested a node-based Susceptible-Latent-Exploding-Susceptible (SLBS) model, and in the same year they presented a heterogeneous node-based SIRS model where each node has the different infected and recovered rates \cite{LX Yang2017}. The above models assume that disease transmission takes place between individuals in networks.

However, diseases are propagated not only by the contact between individuals in the same population, but also by the contact between individuals and infective media. For instance, many human diseases, such as dengue fever, malaria, Chagas disease, and so on, can be transmitted by the infective mosquito. For this case, Shi et al. \cite{Shi2008} established a new SIS epidemic model with an infective medium, which describes epidemics transmitted by infective media on various complex networks. By differentiating the infective medium from individuals, Yang et al. \cite{Meng Yang2011} proposed a modified SIS model. Wang et al. \cite{Wang2012} presented a modified  SIS with an infective vector by incorporating some infectious diseases. It is noteworthy that these existing models with infective media are degree-based instead of node-based.

The motivation of this paper is to build a node-based SIRS epidemic model with infective media on various complex networks by integrating the node-based approach and the infective medium, and investigate the stability of the equilibrium as well as the influence of network structures, the infective medium and the effective recovered rate on the network infected steady state.

The rest of this paper is organized as follows. Some definitions and Lemmas are introduced in Sec. 2. In Sec. 3, a node-based SIRS epidemic network model with infective media is built and then its equilibrium is given. The global asymptotical stability analysis with respect to the equilibrium is performed in Sec. 4. In Sec. 5, numerical simulations of three typical network topologies are provided for further verifying the theoretical results. The correlation between the infected percents of nodes and its degree, as well as the impact of some critical parameters on network average infected percents, are studied theoretically and numerically. Finally, some conclusions and discussions are given in Sec. 6.

\section{Preliminaries}

First, some requisite definitions and lemmas are given as follows.

\textbf{Definition\,1 \cite{Define}:} A Matrix is  \emph{Metzler} if its all off-diagonal entries are non-negative.

\textbf{Definition\,2 \cite{Define}:} A Matrix $A$ is \emph{Hurwitz} stable if there exists a positive matrix $D$ such that $A^TD + DA$ is negative definite.

\textbf{Definition\,3 \cite{Define}:} A Matrix $A$ is diagonally stable if there exists a positive definite diagonal matrix $D$ such that $A^TD + DA$ is negative definite.

Obviously, the diagonally stable matrix is \emph{Hurwitz} stable, and the opposite is also true for \emph{Metzler} matrices.

\textbf{Lemma\,1 \cite{HurwitzMet}:} A \emph{Hurwitz} and \emph{Metzler} matrix is diagonally stable.

\textbf{Lemma\,2 \cite{LX Yang2017}:} Let $A$ be a \emph{Hurwitz} and \emph{Metzler} matrix, $D_1$ be a positive definite diagonal matrix, $D_2$ and $D_3$ be negative definite diagonal matrices. Then,
$$\begin{bmatrix}
A&D_2\\
D_1&D_3
\end{bmatrix}$$
is diagonally stable.


\textbf{Lemma\,3 \cite{compact}:} Consider a smooth dynamical system $\dot{\bm{x}}=g(\bm{x})$ defined at least in a compact set $C$. Then, $C$ is positively invariant if $g(\bm{x}^*)$ is pointing into $C$ for any smooth point $\bm{x}^*$ on the boundary of $C$.

\section{Model formulation}
To begin with, we consider an underlying network( or a simple graph) denoted $G=(V,E)$ where $V$ is the set of nodes and $E$ is the set of edges. The nodes labeled from number $1$ to number $N$ represent the individuals in propagation networks, and the edges stand for network links through which disease can propagate. In a simpler way, we denote $A = (a_{ij})_{N\times N}$ the adjacent matrix of graph $G$ describing network topological structures, where $a_{ij}=1$ if there is an edge between node $i$ and node $j$, otherwise $a_{ij}=0$.

Assume that (H1) each node in the network has three possible states: susceptible($S$), infected ($I$), and recovered ($R$),
whereas the media has two possible states: susceptible ($S^m$) and infected ($I^m$); (H2) both states $S$ and $I$ convert each other with certain probability, and the state $S^m$ is infected with the probability of $\gamma^m$ into the state $I^m$, but not vice versa; (H3) the state $I$ is recovered with the probability of $\lambda_1$ into the state $R$ , and the state $R$ is converted with the probability of $\alpha$ into the state $S$ after the immunity is out of work; (H4) the state $S$ in the underlying network is infected with the probability of $\beta^m$  by the infective media, and the media is with the birth (death) rate of $\mu$.

For simplicity, the variables and parameters in this node-based SIRS model with infective media are summarized in Table 1,
and the schematic diagram of the model is shown in Fig.1.
\begin{figure}[!ht]\label{model}
\centering
  \includegraphics[width=16cm]{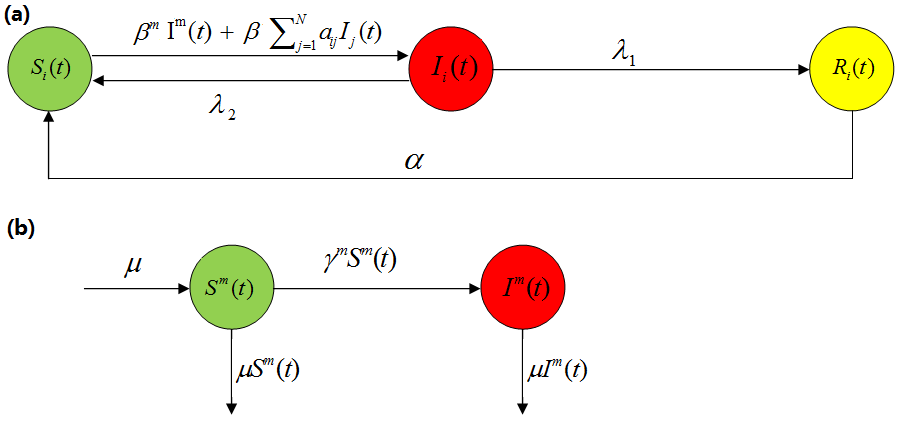}
  \caption{The schematic diagram of node-based SIRS network model with media. (a) The SIRS network part, (b) The media part. }
\end{figure}

\begin{table}
\caption{\newline  Description of parameters.}
\begin{tabular}{c l}
\hline
Parameters & Description\\
\hline
$S_{i}(t)$ & The percents that node i is susceptible at time t.\\
$I_i(t)$ & The percents that node i is infected at time t.\\
$R_i(t)$ & The percents that node i is recovered at time t.\\
$S^m(t)$ & The percents that media is susceptible at time t.\\
$I^m(t)$ & The percents that media is susceptible at time t.\\
$\beta^m$ & The probability that a susceptible node is infected by an infective media.\\
$\beta$ & The probability that a susceptible node is infected by an infected neighbor.\\
$\lambda_1$ & The probability of infective node turns into an immunized one.\\
$\lambda_2$ & The probability of infective node turns into a susceptible one.\\
$\alpha$ & The probability that an immunized node loses immunity into a susceptible one.\\
$\mu$ & The birth (death) rate of the medium.\\
$\gamma^m$ & The probability of a susceptible medium transforming into an infected one.\\
\hline
\end{tabular}
\end{table}

Let $X_i(t)=0$, $1$, and $2$ represent three states of node $i$ at time $t$:
the susceptible ($S_{i}(t)$), the infected ($I_i(t)$) and the recovered ($R_i(t)$), respectively.
$X^m(t)=0$, $1$, and $2$ represent three states of media at time $t$: the susceptible ($S^m(t)$), the infected ($I^m(t)$) and the dead, respectively.
The state of individuals at time t can be expressed as by the vector
    $$X(t)=[X_1(t),X_2(t),...,X_N(t)].$$
Then
$$S_i(t)=P\{X_i(t)=0\},\; I_i(t)=P\{X_i(t)=1\},\; R_i(t)=P\{X_i(t)=2\},$$
$$S^m(t)=P\{X^m(t)=0\},\; I^m(t)=P\{X^m(t)=1\}$$

According to the assumptions, it implies the following probability of state transition:
$$P\{X_i(t+\Delta t)=1|X_i(t)=0\}=\Delta t[\beta^mI^m(t)+\beta\overset{N}{\underset{j =1}{\sum}}a_{ij}I_j(t)]+o(\Delta t)$$
$$P\{X_i(t+\Delta t)=2|X_i(t)=1\}=\lambda_1\Delta t + o(\Delta t)$$
$$P\{X_i(t+\Delta t)=0|X_i(t)=1\}=\lambda_2\Delta t + o(\Delta t)$$
$$P\{X_i(t+\Delta t)=0|X_i(t)=2\}=\alpha\Delta t + o(\Delta t)$$
$$P\{X^m(t+\Delta t)=1|X^m(t)=0\}=\gamma^m\Delta t + o(\Delta t)$$
$$P\{X^m(t+\Delta t)=2|X^m(t)=0\}=\mu\Delta t + o(\Delta t)$$
$$P\{X^m(t+\Delta t)=2|X^m(t)=1\}=\mu\Delta t + o(\Delta t)$$

By using the total probability law, one can obtain
\begin{align*}
I_i(t+\Delta t)=&P\{X_i(t+\Delta t)=1\}\\
               =&P\{X_i(t)=1\}P\{X_i(t+\Delta t)=1|X_i(t)=0\}\\
                &+ P\{X_i(t)=1\}P\{X_i(t+\Delta t)=1|X_i(t)=1\}\\
                &+ P\{X_i(t)=2\}P\{X_i(t+\Delta t)=1|X_i(t)=2\}\\
               =&S_i(t)(1-\Delta t[\beta^mI^m(t)+\beta\overset{N}{\underset{j =1}{\sum}}a_{ij}I_j(t)])\\
                &+I_i(t)\lambda_2\Delta t + R_i(t)\alpha\Delta t + o(\Delta t).
\end{align*}

Let $\Delta t\rightarrow 0$, we get
$$\frac{dS_i(t)}{dt}=-[\beta^mS_i(t)I^m(t)+\beta S_i(t)\overset{N}{\underset{j =1}{\sum}}a_{ij}I_j(t)]+\alpha R_i(t)+\lambda_2I_i(t).$$
Similarly, it is easy to get the equations dominating $I_i(t)$, $R_i(t)$, $R^m(t)$ and $I^m(t)$. Collecting them together,
we have the following $3N+2$ dimensional dynamical system:
%

\begin{equation}\label{model1}
\setlength{\abovedisplayskip}{2pt}
\setlength{\belowdisplayskip}{10pt}
\left\{\begin{array}{l}
\frac{dS_i(t)}{dt}=-[\beta^mS_i(t)I^m(t)+\beta S_i(t)\overset{N}{\underset{j =1}{\sum}}a_{ij}I_j(t)]+\alpha R_i(t)+\lambda_2I_i(t),\\
\frac{dI_i(t)}{dt}=\beta^mS_i(t)I^m(t)+\beta S_i(t)\overset{N}{\underset{j =1}{\sum}}a_{ij}I_j(t)-(\lambda_1+\lambda_2)I_i(t),\\
\frac{dR_i(t)}{dt}=\lambda_1I_i(t)-\alpha R_i(t),\\
\frac{dS^m(t)}{dt}=\mu-\mu S^m(t)-\gamma^m S^m(t),\\
\frac{dI^m(t)}{dt}=\gamma^mS^m(t)-\mu I^m(t),
\end{array}\right.
\end{equation}
with initial condition
$(S_1(0),\cdots,S_N(0),I_i(0),\cdots,I_N(0),R_i(0),\cdots,R_N(0),S^m(0),I^m(0))^T\in \widetilde{\Omega},$
where
$\widetilde{\Omega}=\{(S_1(t),\cdots,S_N(t),I_1(t),\cdots,I_N(t),R_1(t),\cdots,R_N(t),S^m,I^m)^T \in R_{+}^{3N+2} \mid S_i(t)+I_i(t)+R_i(t)=1,S^m+I^m=1, i=1,\cdots,N \}$.

\textbf{Remark 1:} From the view point of continuous-time Markov chain \cite{Markov},
model (\ref{model1}) is an approximation one on account of the linear transition rate instead of exact one from state $S$ to state $I$.
The performance examined in Appendix C shows that model (\ref{model1}) is able to well forecast
the epidemic dynamics of model (\ref{model0}) built by means of Markov chain technique.
Furthermore, the dynamical behaviors of approximation models is more easily studied by applying the stability theory and method,
and thus the similar approximation model is directly built and studied in a large number of related literatures.

Since $S_i(t)+I_i(t)+R_i(t)\equiv1, S^m+I^m\equiv1,\;1\leq i\leq N$,  system (\ref{model1}) can be reduced into the following system:
\begin{equation}\label{model2}
\setlength{\abovedisplayskip}{2pt}
\setlength{\belowdisplayskip}{10pt}
\begin{cases}
\frac{dI_i(t)}{dt}=\beta^m(1-I_i(t)-R_i(t))I^m(t)+\beta(1-I_i(t)-R_i(t))\overset{N}{\underset{j=1}{\sum}}a_{ij}I_j(t)
                    -(\lambda_1+\lambda_2)I_i(t),\\
\frac{dR_i(t)}{dt}=\lambda_1I_i(t)-\alpha R_i(t),\\
\frac{dI^m(t)}{dt}=\gamma^m(1-I^m(t))-\mu I^m(t).
\end{cases}
\end{equation}
with initial condition $(I_1(0),\cdots,I_N(0),R_1(0),\cdots,R_N(0),I^m(0))^T\in \Omega,\ i=1,\cdots,N$,
where $\Omega=\{(I_1(t),\cdots,I_N(t),R_1(t),\cdots,R_N(t),I^m)^T\in R_{+}^{2N+1}|I_i(t)+R_i(t)\leq 1, I^m\leq 1, i=1,\cdots,N \}$.

Let the right-hand terms in (\ref{model2}) equal to zero, one gets an equilibrium $E^*=(I^*_i,R^*_i,{I^m}^*)$ which is only one proven in Appendix A, here
$${I^m}^*=\frac{\gamma^m}{\gamma^m+\mu},\;R^*_i=\frac{\lambda_1}{\alpha}I^*_i,\;
I^*_i=\frac{\frac{\beta^m\gamma^m}{\gamma^m+\mu}+\beta\overset{N}{\underset{j=1}{\sum}}a_{ij}I^*_j}
{\lambda_1+\lambda_2+\beta^m({1+\frac{\lambda_1}{\alpha})\frac{\gamma^m}{\gamma^m+\mu}
}+\beta(1+\frac{\lambda_1}{\alpha})\overset{N}{\underset{j=1}{\sum}}a_{ij}I^*_j}.$$
From the above equality, it is easy to get that $I_i^*<\frac{1}{1+\lambda_1/\alpha}$,
implying that the percent with the infected state for any node is less than $\frac{1}{1+\lambda_1/\alpha}$.

\textbf{Remark 2:} Obviously, the equilibrium is not virus-free, implying that the virus exists persistently in each individual.
The phenomenon can be understood by the fact that the endemic disease remains safely under cover in each individual in some local areas.
Although the equilibrium is given in the implicit form, it can be calculated out by the numerical iterative method.

\textbf{Remark 3:} When the infected rate of media $\beta^m=0$, the model is reduced to the node-based SIRS epidemic model with
 a virus-free equilibrium, to some extent implying that our extended model is rational and practical.
The infected medium terms not only increase the dimension of SIRS models, but more importantly make stability analysis more complicated,
especially in the part of global attractivity.

\section{Stability analysis with respect to equilibria}

\subsection{Local stability}
To analyze the local stability of system (\ref{model2}) at the equilibrium $E^*$, we start with its Jacobian
$$J_{E^*}=\left(
  \begin{array}{cc}
    B & C \\
    \bm{0} & -\gamma^m-\mu \\
  \end{array}
\right),$$
where $C=\beta^m(S_1^*,\cdots,S_N^*,0,\cdots,0)^T$,
$B=\left(
    \begin{array}{cc}
      D &  -\beta^m{I^m}^*E_N - \beta diag(A\bm{I}^*)\\
      \lambda_1E_N & -\alpha E_N \\
    \end{array}
  \right),
$
$\bm{I}^*=(I^*_1,I^*_2,\cdots,I^*_N)^T$, $E_N$ is the identity matrix of order $N$, $diag(\cdot)$ represents the diagonal matrix, and
\begin{equation}\label{Eq-D}
  D=\beta diag(S^*_1,S^*_2,\cdots,S^*_N) A - (\lambda_1+\lambda_2)diag(\frac{S^*_1+I^*_1}{S^*_1},\frac{S^*_2+I^*_2}{S^*_2}, \cdots,\frac{S^*_N+I^*_N}{S^*_N}).
\end{equation}

\textbf{Theorem 1:} System (\ref{model2}) is asymptotically stable at the equilibrium $E^*$

\textbf{Proof:} Obviously, $-\gamma^m-\mu$ is a negative eigenvalue of $J_{E^*}$, and other eigenvalues are determined by matrix $B$. Next, we show that all the eigenvalues of $B$ have negative real part.

For convenience, define three matrices as follows
\begin{align}
K_1=&A - \frac{\lambda_1+\lambda_2}{\beta} diag\left(\frac{1}{S^*_1},\cdots,\frac{1}{S^*_N}\right),\\
K_2=&K_1+\frac{\lambda_1+\lambda_2}{\beta}\max \limits_{i}\left(\frac{1}{S^*_i}\right)E_N,\\
K_3=&A - \frac{\lambda_1+\lambda_2}{\beta}diag\left(\frac{1}{S^*_1},\cdots, \frac{1}{S^*_N}\right) diag\left(\frac{S^*_1+I^*_1}{S^*_1},\cdots,\frac{S^*_N+I^*_N}{S^*_N}\right).
\end{align}
Here we consider an undirected and connected graph, so the adjacent matrix $A$ is an irreducible one,
indicating that $K_2$ is also an irreducible matrix.
According to Perron-Frobenius theorem \cite{Perron}, $K_2$ has a positive eigenvector $\bm{v}$ corresponding to the largest eigenvalue $\lambda_{max}(K_2)$, i.e.,
$K_2\bm{v}=\lambda_{max}(K_2)\bm{v}$. Then
$$
K_1\bm{v}=\left(\lambda_{max}(K_2)-\frac{\lambda_1+\lambda_2}{\beta}\max \limits_{i}\frac{1}{S^*_i}\right)\bm{v},
$$
and
\begin{equation}\label{Eq-K1}
  \bm{v}^TK_1\bm{I}^*=\left(\lambda_{max}(K_2)-\frac{\lambda_1+\lambda_2}{\beta}\max \limits_{i}\frac{1}{S^*_i}\right)\bm{v}^T\bm{I}^*.
\end{equation}

On the other hand, it is easy from the second equality of system (1) to get that
$$\beta^m{I^m}^*S^*+\beta diag(S^*)A\bm{I}^*-(\lambda_1+\lambda_2)\bm{I}^*=0, $$
where $S^*=(S^*_1,\cdots,S^*_N)^T$. Thus, $(\beta diag(S^*))K_1\bm{I}^*=-\beta^m{I^m}^*S^*\leq0$ implies $K_1\bm{I}^*\leq 0$ when $\beta S_1^*S_2^*\cdots S_N^*\neq 0$.

Since $\bm{v}$ and $\bm{I}^*$ are positive vectors, it holds that
$$\lambda_{max}(K_2)-\frac{\lambda_1+\lambda_2}{\beta}\max \limits_{i}\frac{1}{S^*_i}\leq 0,$$
and
$$\lambda_{max}(K_1)=\lambda_{max}(K_2)-\frac{\lambda_1+\lambda_2}{\beta}\max \limits_{i}\frac{1}{S^*_i}\leq 0.$$

From (6) and (7), it follows that
\begin{equation*}
\lambda_{max}(K_3)<\lambda_{max}\left(A - \frac{\lambda_1+\lambda_2}{\beta}diag(\frac{1}{S^*_1},\cdots, \frac{1}{S^*_N})\right)=\lambda_{max}(K_1)\leq 0.
\end{equation*}
\vskip 0.2cm \noindent
Thus, $K_3$ is a negative definite matrix, and then $D=\beta diag(S^*)K_3$ is also a negative definite one. That is to say, $D$ is a \emph{Hurwitz} and \emph{Metzler} matrix. According to Lemma 2, matrix $B$ is diagonally stable. Therefore, the equilibrium $E^*$ of system (\ref{model2}) is asymptotically stable. $\hfill{} \Box$

\subsection{Global attractivity}
To proof the global attractivity, it needs to determine the positively invariant set (In brief, once a trajectory of the system enters the set, it will never leave it again). Next, it is not difficult to proof that $\Omega=\{(I_1(t),\cdots,I_N(t),R_1(t),\cdots,R_N(t),I^m)^T|I_i(t)+R_i(t)\leq 1, I^m\leq 1, i=1,\cdots,N \}$ is an invariant set.

\textbf{Theorem\,2:} $\Omega$ is a positively invariant set for system (2).

\textbf{Proof:} Denote $\partial\Omega$ the boundary of $\Omega$, and then it consists of the following $3N+2$ hyperplanes:
\begin{align*}
&\Gamma_i=\{(I_1(t),\cdots,I_N(t),R_1(t),\cdots,R_N(t),I^m(t))^T\in \Omega|I_i(t)=0\},\;i=1,\cdots,N,\\
&\Gamma_{N+i}=\{(I_1(t),\cdots,I_N(t),R_1(t),\cdots,R_N(t),I^m(t))^T\in \Omega|R_i(t)=0\},\;i=1,\cdots,N,\\
&\Gamma_{2N+i}=\{(I_1(t),\cdots,I_N(t),R_1(t),\cdots,R_N(t),I^m(t))^T\in \Omega|I_i(t)+R_i(t)=1\},\;i=1,\cdots,N,\\
&\Gamma_{3N+1}=\{(I_1(t),\cdots,I_N(t),R_1(t),\cdots,R_N(t),I^m(t))^T\in \Omega|I^m(t)=0\},\\
&\Gamma_{3N+2}=\{(I_1(t),\cdots,I_N(t),R_1(t),\cdots,R_N(t),I^m(t))^T\in \Omega|I^m(t)=1\},
\end{align*}

For simplicity and convenience, system (\ref{model2}) is rewritten as:
$$\frac{dz(t)}{dt}=g(z(t))$$
with initial condition $z(0)\in\partial\Omega$.

Take the outer normal vectors corresponding to $3N+2$ hyperplanes as follows:
\begin{align*}
&P_i=(\underbrace{0,\cdots,-1}_i,\cdots,0,0,\cdots,0)^T,\\
&P_{N+i}=(\underbrace{0,\cdots,0,0,\cdots,-1}_{N+i},\cdots,0)^T,\\
&P_{2N+i}=(\underbrace{\underbrace{0,\cdots,1}_i,\cdots,0,0,\cdots,1}_{N+i},\cdots,0)^T,\\
&P_{3N+1}=(\underbrace{0,\cdots,0,0,\cdots,0,-1}_{2N+1})^T,\\
&P_{3N+2}=(\underbrace{0,\cdots,0,0,\cdots,0,1}_{2N+1})^T,
\end{align*}
and let $z^*=(I^*_1,\cdots,I^*_N,R^*_1,\cdots,R^*_N,{I^m}^*)^T$ be a smooth point of $\partial\Omega$.
On the basis of these hyperplanes, five different cases of $z^*$ are discussed respectively.
\begin{itemize}
  \item Case 1: For $I^*_i=0$, $(\frac{dz}{dt}|_{z^*\in {\Gamma}_i},P_{i})=
                                     -\beta^m(1-R^*_i){I^m}^*-\beta(1-R^*_i)\overset{N}{\underset{j=1}{\sum}}a_{ij}I^*_j<0.$
  \item Case 2: For $R^*_i=0$, $(\frac{dz}{dt}|_{z^*\in {\Gamma}_{N+i}},P_{N+i})= -\lambda_1I^*_i<0.$
  \item Case 3: For $I^*_i+R^*_i=1$, $(\frac{dz}{dt}|_{z^*\in {\Gamma}_{2N+i}},P_{2N+i})= -\lambda_2I^*_i-\alpha R^*_i<0.$
  \item Case 4: For ${I^m}^*=0$, $(\frac{dz}{dt}|_{z^*\in {\Gamma}_{3N+1}},P_{3N+1})= -\gamma^m<0.$
  \item Case 5: For ${I^m}^*=1$, $(\frac{dz}{dt}|_{z^*\in {\Gamma}_{3N+2}},P_{3N+2})= -\mu<0.$
\end{itemize}
Therefore, $g(z^*)$ is pointing to $\Omega$, and $\Omega$ is positively invariant according to Lemma\,3.   $\hfill{} \Box$

\textbf{Theorem 3:} The equilibrium $E^*$ of system (2) is globally attractive on $\Omega-\{0\}$.

\textbf{Proof:} Denote $\bm{y}(t)=(y_1(t),\,\cdots,y_{2N}(t),\,y_{2N+1}(t))^T$ where
$y_i(t)=I_i(t),\,y_{N+i}=R_i(t),\,y_{2N+1}=I^m(t) \,(i=1,2,\cdots,N)$,
and then the virus equilibrium $\bm{y}^*=(y_1^*,\cdots,y_{2N+1}^*)$, here $y_i*\neq0\,(i=1,\cdots,N)$.

To explore the asymptotic behavior of solutions of Eq. (\ref{model2}), we define two functions as follows:
\begin{align*}
F(\bm{y}(t))&=\underset{i}{max}{\frac{y_i(t)}{y^*_i}}:\Omega\rightarrow R,\\
f(\bm{y}(t))&=\underset{i}{min}{\frac{y_i(t)}{y^*_i}}:\Omega\rightarrow R.
\end{align*}

\noindent
Both functions are continuous and exist right-hand derivatives along solutions of Eq.(\ref{model2}).
Let $\bm{y}(t)$ is the solution of Eq.(\ref{model2}), and suppose that $F(\bm{y}(t))=\frac{y_{i_0}(t)}{y_{i_0}^*}, t\in [t_0,t_0+\varepsilon]$, for some $t_0$ and sufficiently small $\varepsilon >0$. Then we have

$$F'(\bm{y}(t_0))|_{(2)}=\frac{y'_{i_0}(t_0)}{y^*_{i_0}},$$
where $F'(\bm{y}(t))|_{(2)}\triangleq\underset{h\to 0^+}{\lim}{sup}\frac{F(\bm{y}(t+h))-F(\bm{y}(t))}{h}$.

Next, we proof that the derivative of $F(\bm{y}(t))$ at $t_0$ is non-negative.
According to the definition of $F(\bm{y}(t))$, it follows that
$$\frac{y_{i_0}(t_0)}{y_{i_0}^*}\geq\frac{y_i(t_0)}{y_i^*},\,i=1,2,\cdots,2N+1.$$
For $F(\bm{y}(t))>1$, three cases as below are discussed (here $t_0$ is ignored for conciseness).
\begin{itemize}
  \item Case 1: $1\leq i_0\leq N.$
\begin{align*}
y^*_{i_0}\frac{y'_{i_0}(t_0)}{y_{i_0}(t_0)}=&\frac{y^*_{i_0}}{y_{i_0}}
\{\beta^m(1-y_{i_0}-y_{N+i_0})y_{2N+1}+\beta (1-y_{i_0}-y_{N+i_0})\overset{N}{\underset{j =1}{\sum}}a_{ij}y_j-(\lambda_1+\lambda_2)y_{i_0}\}\\
< &\beta^m(1-y^*_{i_0}-y^*_{N+i_0})y_{2N+1}\frac{y^*_{i_0}}{y_{i_0}}+\beta (1-y^*_{i_0}-y^*_{N+i_0})
\frac{y^*_{i_0}}{y_{i_0}}\overset{N}{\underset{j =1}{\sum}}a_{ij}y_j-(\lambda_1+\lambda_2)y^*_{i_0}\\
<
&\beta^m(1-y^*_{i_0}-y^*_{N+i_0})y^*_{2N+1}+\beta(1-y^*_{i_0}-y^*_{N+i_0})
 \overset{N}{\underset{j =1}{\sum}}a_{ij}y^*_j-(\lambda_1+\lambda_2)y^*_{i_0}=0.
\end{align*}

  \item Case 2: $N+1\leq i_0\leq 2N$. $y^*_{i_0}\frac{y'_{i_0}(t_0)}{y_{i_0}(t_0)}=(\lambda_1y_{i_0-N}-\alpha y_{i_0})
\frac{y^*_{i_0}}{y_{i_0}}<\lambda_1y^*_{i_0-N}-\alpha y^*_{i_0}=0$.

  \item Case 3: $i_0=2N+1$. $y^*_{i_0}\frac{y'_{i_0}(t_0)}{y_{i_0}(t_0)}=(\lambda_1y_{i_0-N}-\alpha y_{i_0})
\frac{y^*_{i_0}}{y_{i_0}}<\lambda_1y^*_{i_0-N}-\alpha y^*_{i_0}=0.$

\end{itemize}
Since $y_{i_0}^*>0$ and $y_{t_0}(t_0)>0$, one get $y'_{i_0}(t_0)<0$, implying $F'(\bm{y}(t_0))<0$.
Similarly, $F'|_{(2)}(\bm{y}(t_0))\leq 0$ if $F(\bm{y}(t_0))=1$, $f'|_{(2)}(\bm{y}(t_0))>0$ if $f(\bm{y}(t_0))<1$, and $f'|_{(2)}(\bm{y}(t_0)\geq 0$ if $f(\bm{y}(t_0))=1$.

Denote
\begin{align*}
U(\bm{y})&=max\{F(\bm{y})-1,0\},\;\bm{y}\in\Omega,\\
V(\bm{y})&=min\{1-f(\bm{y}),0\},\;\bm{y}\in\Omega.
\end{align*}
Obviously, $U(\bm{y})$ and $V(\bm{y})$ are non-negative and continuous in $\Omega$,  and $U'(\bm{y})|_{(5)}\leq0$ and $V'(\bm{y})|_{(5)}\leq0$.
Let $H_U=\{\bm{y}\in \Omega|U'|_{(2)}(\bm{y}(t))=0\}$ and $H_V=\{\bm{y}\in \Omega|V'|_{(2)}(\bm{y}(t))=0\}$, then we have
$H_U=\{\bm{y}:0\leq y_j(t)\leq y^*_j\}\cup\{0\}$ and $H_V=\{\bm{y}:y^*_j\leq y_j(t)\leq 1\}\cup\{0\}$.

It follows from the LaSalle Invariance Principle that any solution of system (\ref{model2}) staring in $\Omega$ approaches $H_U\cap H_V=\{\bm{y}^*\}\cup\{0\}$. Therefore, any solution $\bm{y}(t)$ with initial value $\bm{y}(0)\in \Omega$ satisfies
$\lim_{t\rightarrow\infty}\bm{y}(t)=\bm{y}^*$, i.e., $\bm{y}^*$ is globally attractive in $\Omega$.

\textbf{Remark 4:} Together with local asymptotical stability, it is easily obtained that the equilibrium $E^*$ in the SIRS model with media is globally asymptotically stable.  By the way, without the medium propagation, i.e., $\beta^m=0$, system (\ref{model2}) has a virus-free equilibrium $E_0$
which is globally asymptotically stable if $\sigma_{max}(A)<(\lambda_1+\lambda_2)/\beta$, otherwise unstable.
Here $\sigma_{max}(A)$ is the largest eigenvalue of the topological matrix $A$, and $(\lambda_1+\lambda_2)/\beta$ represents the actual effective recovered rate.

\section{Numerical simulations}
\subsection{Three typical network models}
In order to verify the above theoretical results, we choose three typical network structures (fully connected network, small-world network and scale-free network), and solve numerically system (2) with $\beta^m=0.02,\beta=0.06,\lambda_1=0.3,\lambda_2=0.3,\alpha=0.49,\mu=0.5,\gamma^m=0.15$.
Here, the fully connected network means that all nodes are connected with each other, the small-world network is generated from nearest neighbor network with the probability $p=0.05$ of random adding edges (called NW small-world network), and the scale-free network is a BA scale-free one with $m_0=10$ and $m=3$. Each type of network is with 100 nodes, respectively.

\begin{figure}[!ht]
\centering
\includegraphics[scale=0.4]{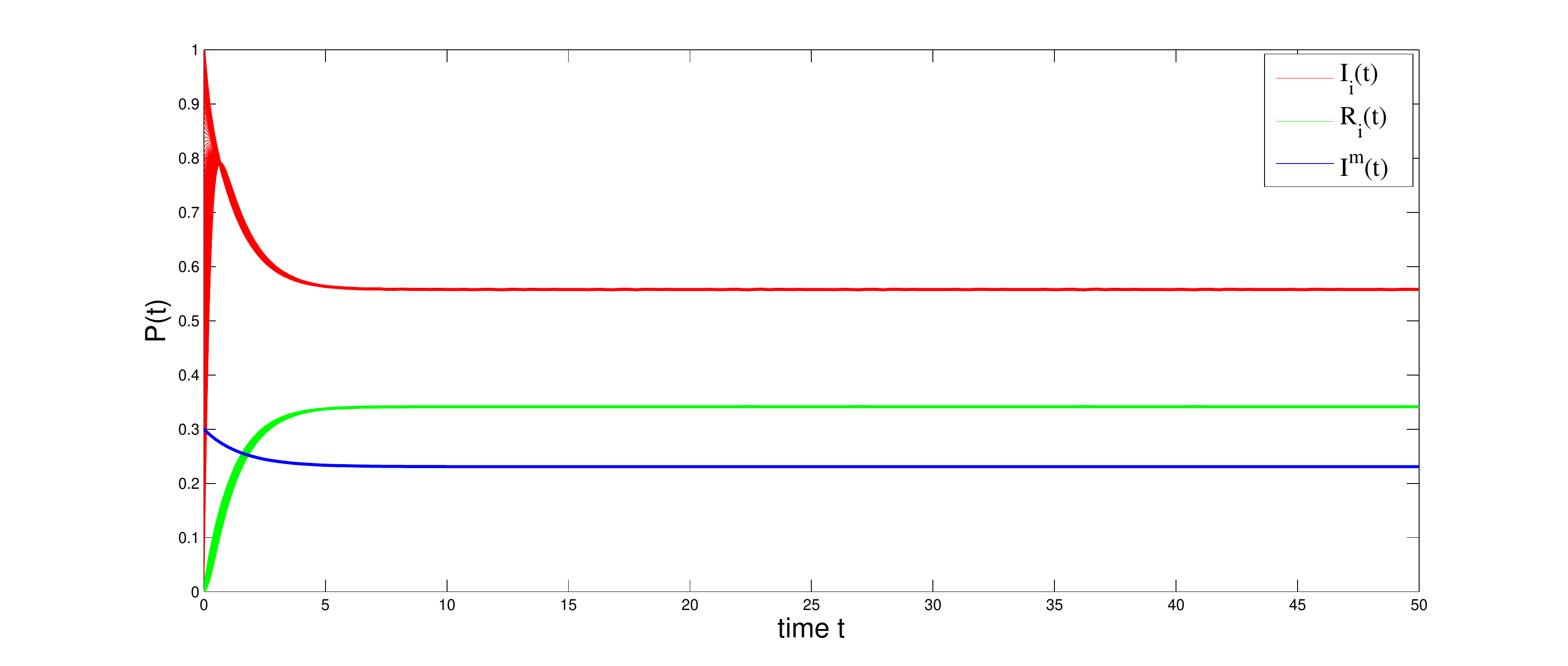}
\caption{\small The states $I_i(t)$, $R_i(t)$ and $I^m(t)$ of time for fully connected networks and random initial conditions.}
\label{fig:shuzhicomplete}
\end{figure}

\begin{figure}[!ht]
\centering
\includegraphics[scale=0.4]{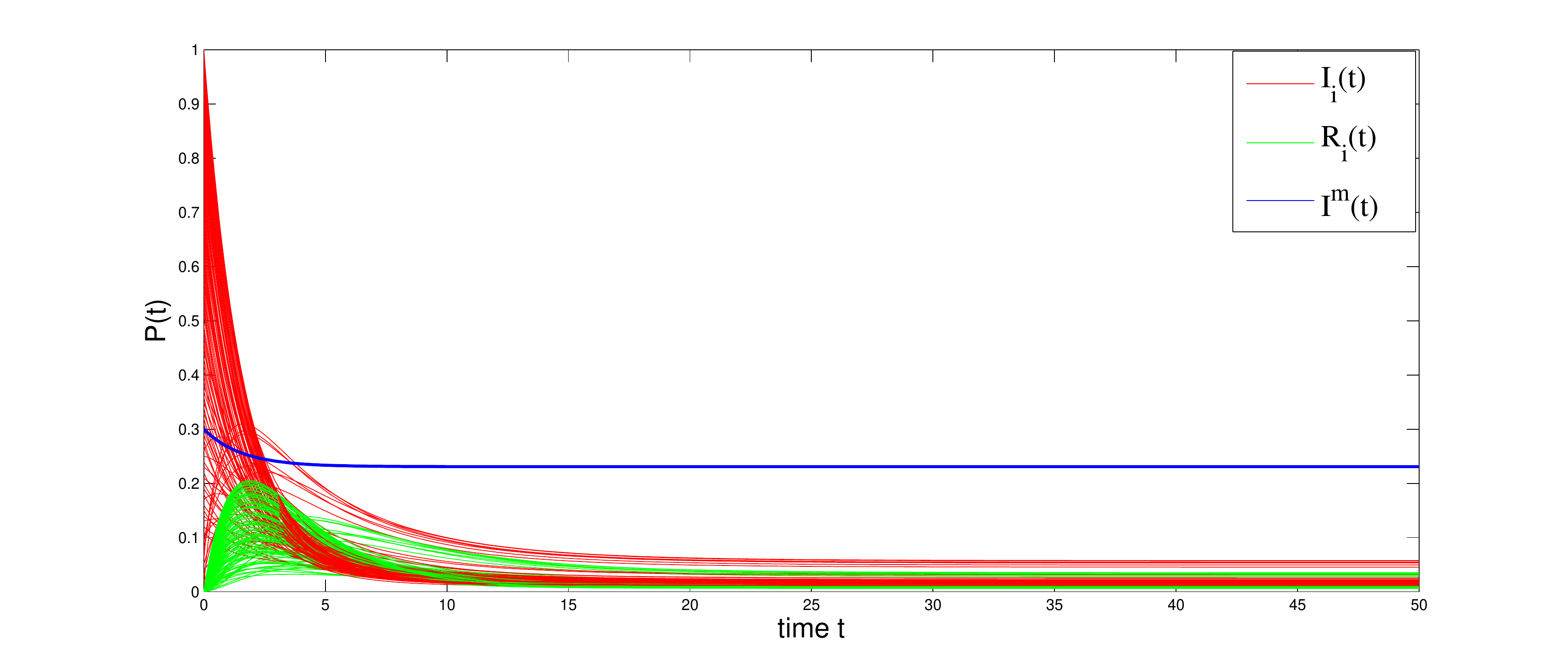}
\caption{\small The states $I_i(t)$, $R_i(t)$ and $I^m(t)$ of time for NW small-world networks and random initial conditions.}
\label{fig:shuzhiNW}
\end{figure}

\begin{figure}[!ht]
\centering
\includegraphics[scale=0.4]{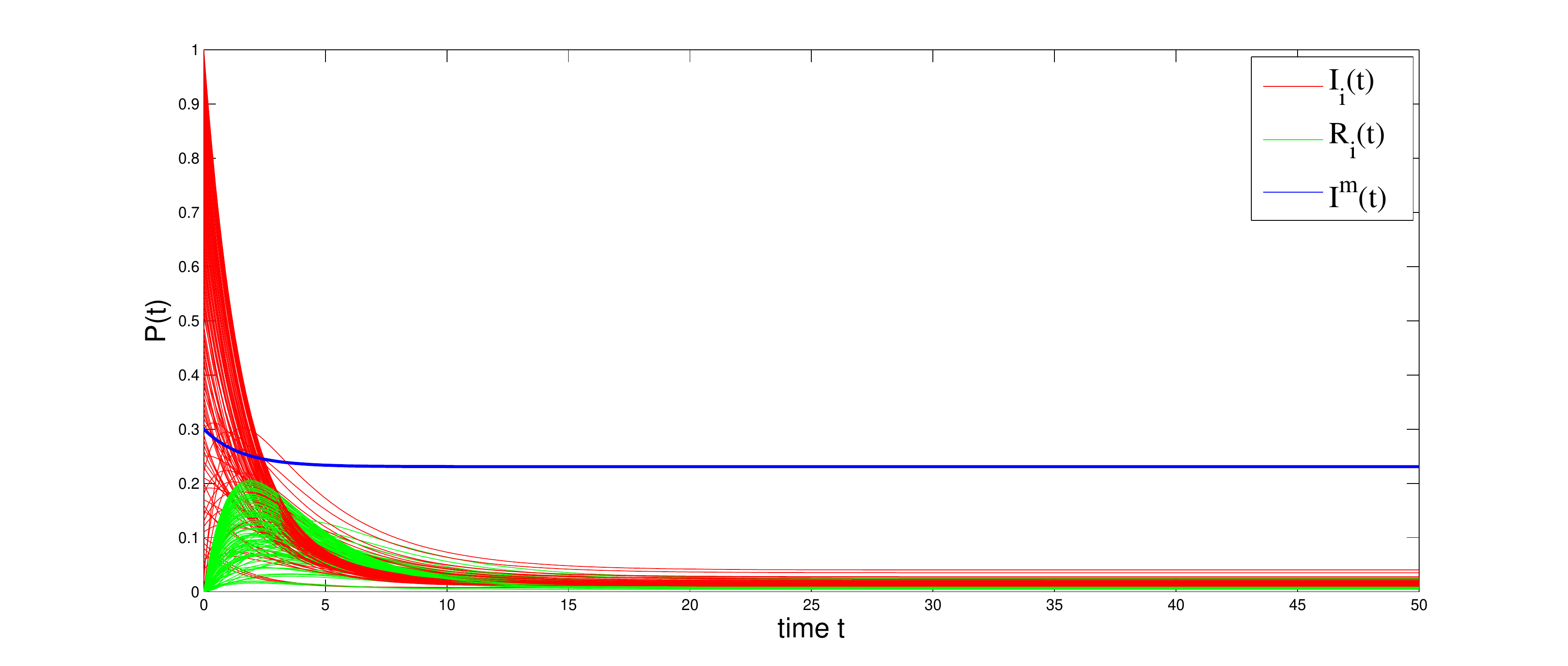}
\caption{\small The states $I_i(t)$, $R_i(t)$ and $I^m(t)$ of time for BA scale-free networks and random initial conditions.}
\label{fig:shuzhiSF}
\end{figure}

Figures \ref{fig:shuzhicomplete}-\ref{fig:shuzhiSF} show the evolution of $I_i(t)$, $R_i(t)$ and $I^m(t)$ over time for the fully connected network, NW small-world network and BA scale-free network, respectively. Obviously, the infected ($I$) and susceptible ($S$) states of each node in each network tends to their equilibrium states over time, namely, $I_i(t)\rightarrow I_i^*$ and $S_i(t)\rightarrow S_i^*$($t\rightarrow\infty$), implying that it further verifies our theoretical results, and the infected percents of node is very low for sparsely connected networks, such as NW and BA networks.

Interestingly, for the fully connected network, all nodes tend to the same equilibrium state as time goes, namely, $I_i(t)\rightarrow I^*$ and $S_i(t)\rightarrow S^*$($t\rightarrow\infty$). But for the small-world network and scale-free network, all nodes approach their equilibrium but nonidentical states as time goes, namely, $I_i(t)\rightarrow I_i^*$ and $S_i(t)\rightarrow S_i^*$($t\rightarrow\infty$). In fact, the steady state of each node is closely related with its degree according to the formula of equilibria. For the fully connected network, each node has the same degree, resulting into that each node tends to the same equilibrium state, and the equilibrium states $I^*$ satisfies   
$$
-\frac{\lambda_1+\lambda_2}{\beta(1+\frac{\lambda_1}{\alpha})(N-1)}<I^*-\frac{1}{1+\frac{\lambda_1}{\alpha}}<-\frac{\lambda_1+\lambda_2}{2\beta(1+\frac{\lambda_1}{\alpha})(N-1)},
$$
and thus $I^*\approx\frac{1}{1+\lambda_1/\alpha}$ for enough large size networks.
Please refer to Appendix B for the detailed derivation.

\begin{figure}[!ht]
\centering
\includegraphics[scale=0.45]{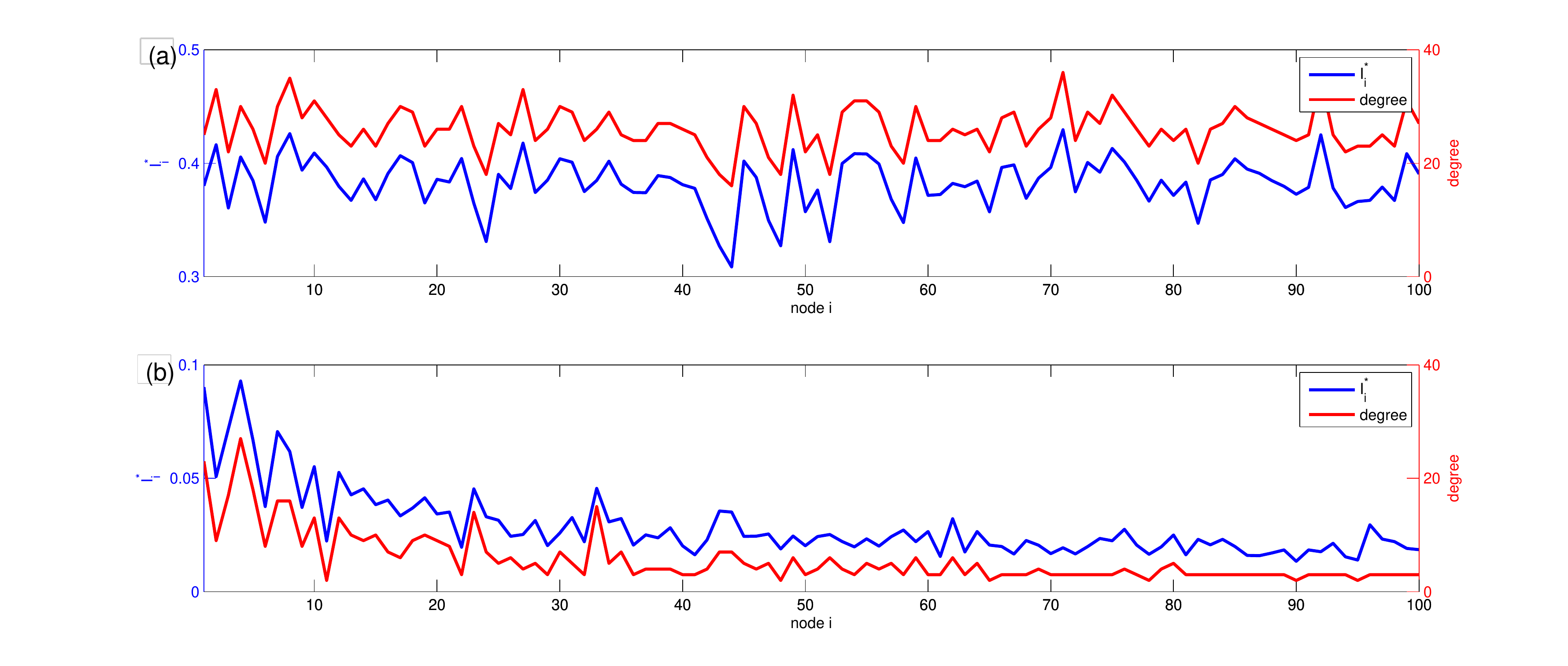}
\caption{\small The correlation between the node degree (red solid line) and its infected equilibrium state (blue solid line) for NW small-world (a) and BA scale-free networks (b).}
\label{fig:degree}
\end{figure}

For the NW small-world network, the node with large (small) degree has large (small) infected equilibrium state, as shown in Fig. \ref{fig:degree} (a). However, the correlation is a little weaker for the BA scale-free network which is a heterogeneous one, please see Fig. \ref{fig:degree} (b).
In a word, the infected (susceptible) equilibrium state $I_i^*$ ($S_i^*$) is positively (negatively) correlated with the degree of nodes, and the node with higher degree is easier to be infected, further verifying the result obtained by the article \cite{yangan2005}.

\bigskip

\subsection{Impact of system parameters}
To learn more about this proposed model, we now analyze the influence of several important parameters, namely, $\beta^m$, $\beta$, $\lambda_1$, and $\alpha$, on network average infected state.

Let us look back these parameters, $\beta^m$ means the infected rate resulted from the medium, $\beta$ represents the infected rate from the node's neighbour, and $\alpha$ can be considered as the potential infected rate due to the fact that the state of nodes never transforms back once it changes to the susceptible state from the recovered state, and the susceptible state is the potential part transferred into the infected state. $\lambda_1$ represents the recovered rate from the infected state to the recovered state.
Here we call $\lambda_1/\beta$ and $\lambda_1/\alpha$ the actual and potential effective recovered rate, respectively.

We firstly analyze the influence of $\beta^m$, $\alpha$, $\beta$ and $\lambda_1$ on network average infected state for the general topologies.
According to the implicit differentiation theorem, it is easy to obtain the following theoretical results through the formula of equilibria.

\textbf{Theorem\,3:} Under the assumptions (H1)-(H4), it follows that
(a) $\frac{\partial \bar{I}^*}{\partial \beta^m}>0,$
(b) $\frac{\partial \bar{I}^*}{\partial \alpha}>0,$
(c) $\frac{\partial \bar{I}^*}{\partial \beta}>0,$
(d) $\frac{\partial \bar{I}^*}{\partial \lambda_1}<0,$
where $\bar{I}^*=\frac{1}{N}\sum_{i=1}^N I_i^*$ called network average infected state. (Please see Appendix D for the proof)

The above theorem implies that the network average infected state rises with the increase of $\beta^m$,
and decreases with the increase of $\lambda_1/\alpha$ or $\lambda_1/\beta$.

\begin{figure}[!ht]
\centering
\includegraphics[scale=0.6]{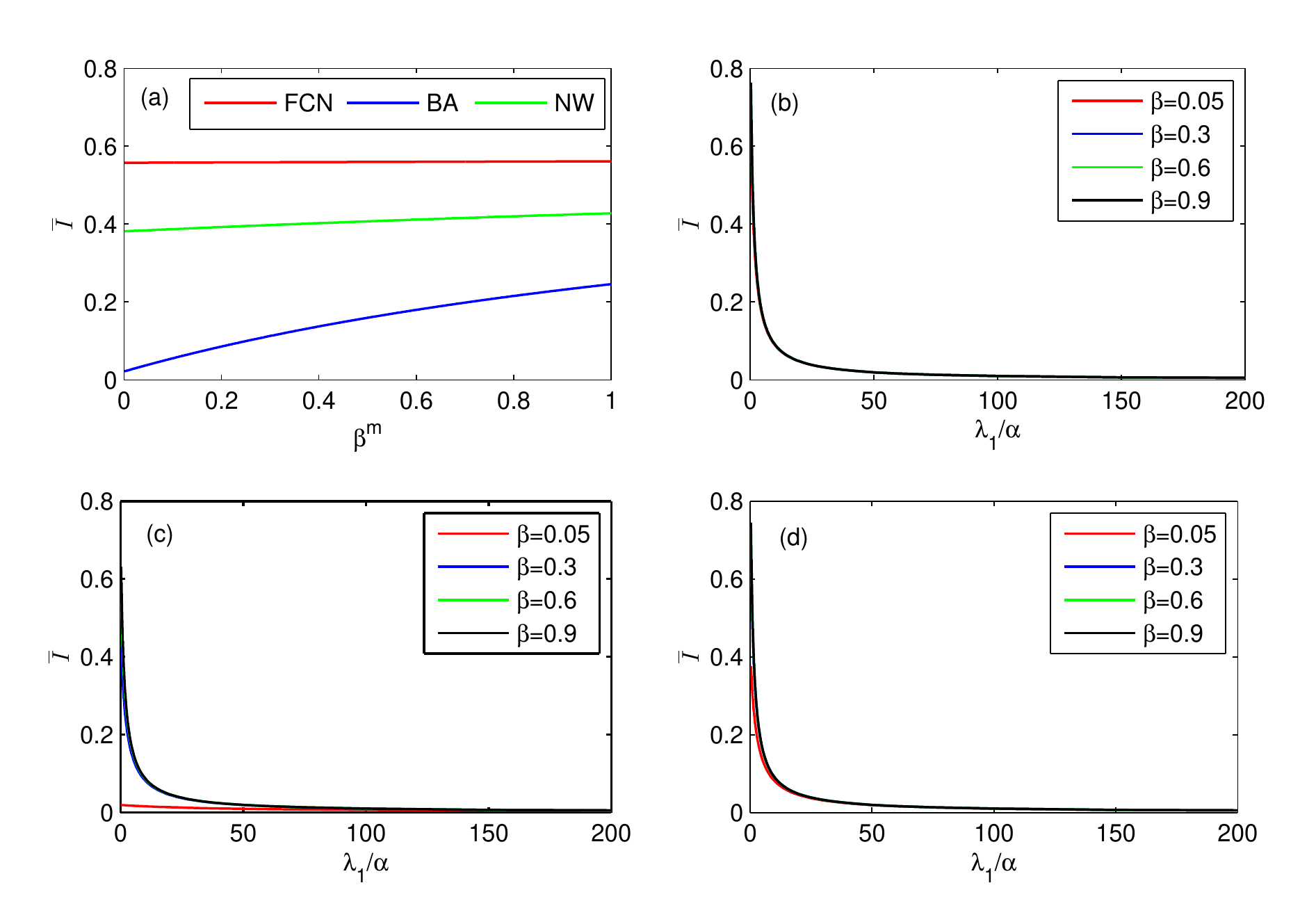}
\caption{\small (a) The influence of the infected rate $\beta^m$ of media on network average infected states, and the curves from top to bottom correspond to the fully-connected network, NW small-world network and BA scale-free network, respectively. The influence of the potential spreading rate $\lambda_1/\alpha$ on network average virus at different values of $\beta$ for the fully-connect network (b), NW small-world network (c) and BA scale-free networks (d). The curves from bottom to top in each subplot correspond to the value of $\beta=0.05,\,0.3,\, 0.6,\, 0.9$, respectively.}
\label{fig:beta}
\end{figure}

Furthermore, we numerically verify the theoretical implications for three typical network topologies, respectively. Fig. \ref{fig:beta}(a) shows that with fixed $\lambda_1=0.3, \lambda_2=0.3, \beta=0.06, \alpha=0.49, \gamma^m=0.15, \mu=0.5$,  as $\beta^m$ increases, the network average infected state rises gradually for BA scale-free network, but almost unchanged for the fully-connected network and NW small-world network, implying that the heterogenous network is sensitive to the infected rate of the medium, but the homogeneous network is the opposite.

It is shown from Fig. \ref{fig:beta}(c)(b)(d) that the average infected state of three typical networks decreases exponentially as the potential recovered rate $\lambda_1/\alpha$ increases with fixed $\alpha=0.3$, and it can be located at low level when $\lambda_1/\alpha$ is enough large. It implies that the epidemic spreading on networks can be significantly suppressed by the even small increase of the potential effective recovered rate $\lambda_1/\alpha$.

\begin{figure}[!ht]
\centering
\includegraphics[scale=0.6]{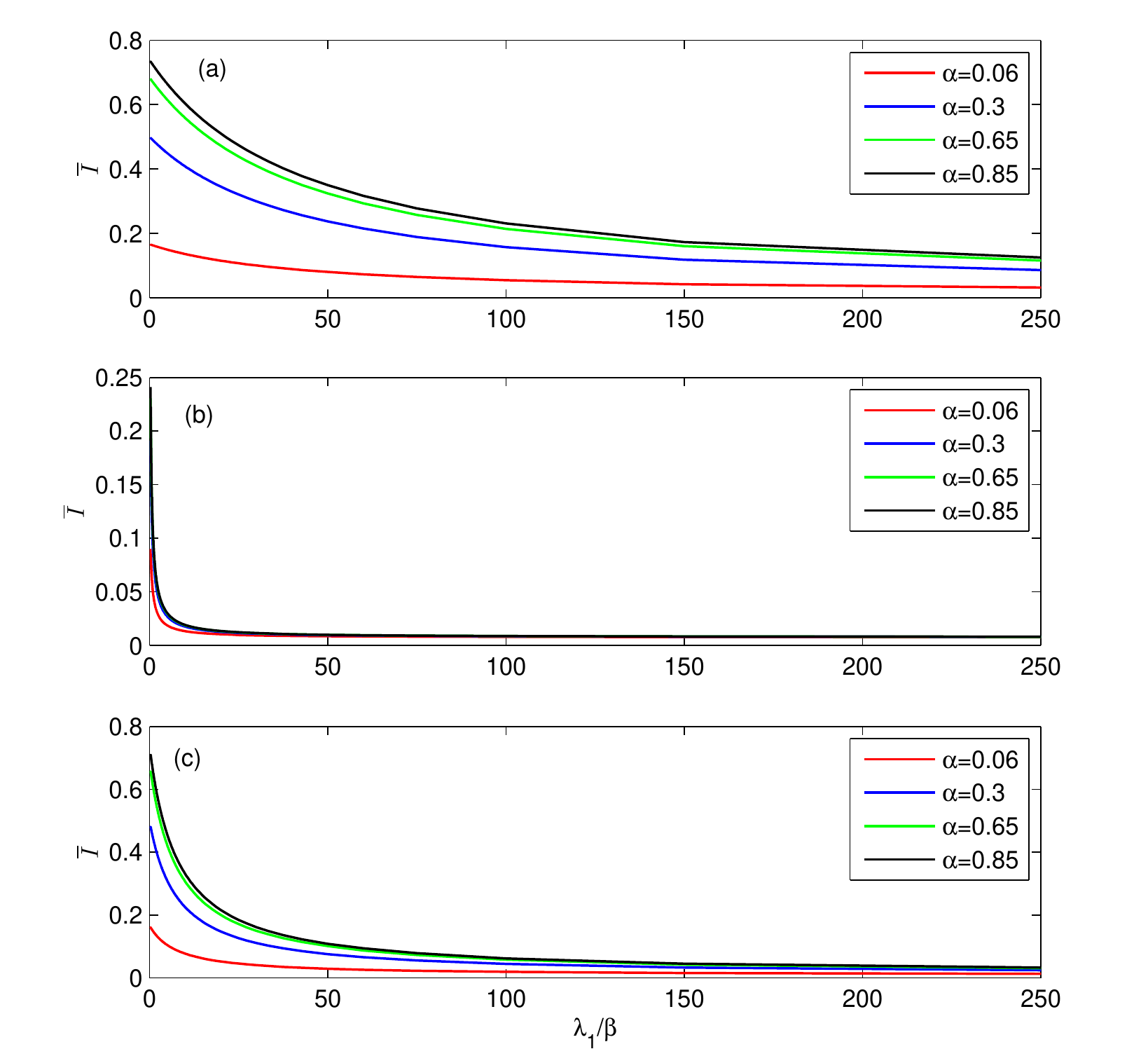}
\caption{\small The influence of the actual spreading rate $\lambda_1/\beta$ on network average infected steady state at different values of $\alpha$ for the fully-connect network (a), NW small-world network (b) and BA scale-free networks (c). The curves from bottom to top in each subplot correspond to the value of $\alpha=0.06,\,0.3,\, 0.65,\, 0.85$, respectively.}
\label{fig:alpha}
\end{figure}

Similarly, the average infected state goes down with the increase of $\lambda_1/\beta$, and the virus decreases exponentially for NW small-world network, faster than those for the fully-connected network and BA scale-free network, as shown in Fig. \ref{fig:alpha}. A possible reason is that the small-world network has the properties of the short average path length and small average degree.

\section{Conclusions and discussions}
In summary, this paper has presented a node-based SIRS epidemic model with media for understanding the disease spreading of networks with media propagation, where there is only an equilibrium yet not virus-free one that is always globally asymptotically stable through the stability analysis.
Without the medium propagation, the model has a virus-free equilibrium which is globally asymptotically stable when the maximum eigenvalue of topological matrices is less than the effective recovered rate $(\lambda_1+\lambda_2)/\beta$.

Three typical networks, i.e., the fully-connected, small-world, and scale-free networks, are applied to numerical investigations for further verifying the theoretical results. numerical simulations also show that the sparse network has less infected percents. In addition, it shows that the infected percents of network nodes have the positive correlation with the degree of the node, in particular for the homogenous network, such as the fully-connected network and small-world network.

Finally, theoretical and numerical studies on the influence of the effective recovered rate and medium propagation rate on network average infected percents imply that network average infected percents go up (down) with the increase of the medium propagation rate (the effective recovered rate). Numerical investigations further show that the medium propagation rate does nothing with network average infected percents for homogenous networks, and the infected percents decease exponentially with the increase of the effective recovered rate. Moreover, the percents can be controlled at low level only if the effective recovered rate is enough large, in other words, only if the effective infected rate is enough small.

\section*{Conflicts of Interest}
The authors declare that they have no conflicts of interest regarding the publication of this paper.

\section*{Acknowledgements}
This work is supported in part by the National Natural Science Foundation of China (Grants No. 61573004, 11871231 and 11501221), in part by the Promotion Program for Young and Middle-aged Teacher in Science and Technology Research of Huaqiao University (ZQN-YX301),
in part by the Program for New Century Excellent Talents in Fujian Province University in 2016, and in part the Project of Education and Scientific Research for Middle and Young Teachers in Fujian Province(JAT170027, JA15030).

\section*{Appendix A: Proof of uniqueness of equilibria}
Here we now prove the equilibrium $E^*$ is one and only equilibrium fixed point.
First of all,
Define a continuous mapping $H=(h_1,\cdots,h_N):(0,\infty)^N\to(0,1)^N$ as below:
\begin{align*}
h_i(\bm{y})=\frac{\frac{\beta^m\gamma^m}{\gamma^m+\mu}+\beta\overset{N}{\underset{j=1}{\sum}}a_{ij}y_j}
{\lambda_1+\lambda_2+\beta^m({1+\frac{\lambda_1}{\alpha})\frac{\gamma^m}{\gamma^m+\mu}
}+\beta(1+\frac{\lambda_1}{\alpha})\overset{N}{\underset{j=1}{\sum}}a_{ij}y_j}
,\quad i=1,\cdots,N.
\end{align*}
We can assert that the equilibrium is one and only if $H$ is monotonic and exist a unique fixed point.

\textbf{Claim\,1:} $H$ is monotonic.

\textbf{Proof:} Let $\bm{X}, \bm{Z} \in (0,\infty)^N$, $\bm{X}\leq \bm{Z},$ ($X_i\leq Z_i,\; i=1,\cdots,N).$ Then,
\begin{align*}
h_i(\bm{X})&=\frac{\frac{\beta^m\gamma^m}{\gamma^m+\mu}+\beta\overset{N}{\underset{j=1}{\sum}}a_{ij}X_j}
{\lambda_1+\lambda_2+\beta^m(1+\frac{\lambda_1}{\alpha})\frac{\gamma^m}{\gamma^m+\mu}
+\beta(1+\frac{\lambda_1}{\alpha})\overset{N}{\underset{j=1}{\sum}}a_{ij}X_j}\\
&\leq \frac{\frac{\beta^m\gamma^m}{\gamma^m+\mu}+\beta\overset{N}{\underset{j=1}{\sum}}a_{ij}Z_j}
{\lambda_1+\lambda_2+\beta^m(1+\frac{\lambda_1}{\alpha})\frac{\gamma^m}{\gamma^m+\mu}
+\beta(1+\frac{\lambda_1}{\alpha})\overset{N}{\underset{j=1}{\sum}}a_{ij}Z_j}=h_i(\bm{Z}),
\end{align*}
which implies $H(\bm{X})\leq H(\bm{Z})$, the proof of Claim 1 is completed.          $\hfill{} \Box$   \\

\textbf{Claim\,2:} $H$ admits a unique fixed point in $(0,1)^N$.

\textbf{Proof:} (1) Existence. Since $H(\bm{y})$ is monotonic and continue for $\bm{y}\in[0,\infty)^N$, it follows that $h_i(0)<h_i(\bm{\xi})<h_i(1),\,\forall\,\bm{\xi}\in(0,1)^N$. On the other hand, $h_i(0)>0$ and $h_i(1)<1$,

implying $\exists\,\eta\in(0,h_i(0))^N$ and $\exists\,\mu\in(h_i(1),1)^N$,
such that $\forall\,\zeta\in[\eta,\mu]^N\subseteq(0,1)^N$ where
$\eta\triangleq(\eta_1,\cdots,\eta_N)$ and $\mu\triangleq(\mu_1,\cdots,\mu_N)$.
we conclude that $\eta_i\leq h_i(\zeta)\leq \mu_i,$
$\eta_i<\mu_i, i=1,\cdots,N,$ so the restriction $H$ on the compact convex set
$$\Lambda=[\eta_1,\mu_1]\cdot[\eta_2,\mu_2]\cdots[\eta_N,\mu_N].$$
maps $\Lambda$ into $\Lambda$. It follows from \emph{Brouwer Fixed point Theorem}\cite{Brower} that H exists a fixed point $U^*\in\Lambda$.

(2) Uniqueness. Suppose $H$ exists the other fixed point $V^*=(v^*_1,\cdots,v^*_N)^T \in (0,1)^N$.
Let $\tau=\underset{i}{\max}\frac{U^*_i}{V^*_i}$ and $i_0=arg\underset{i}{max}\frac{U^*_i}{V^*_i},$ Without loss of generality, we may assume $\tau>1,$ it follows that
\begin{align*}
U^*_{i_0}&=h_{i_0}(U^*)\leq h_{i_0}(\tau V^*)\\
&=\frac{\frac{\beta^m\gamma^m}{\gamma^m+\mu}+\tau\beta\overset{N}{\underset{j=1}{\sum}}a_{ij}V^*_j}
{\lambda_1+\lambda_2+\beta^m(1+\frac{\lambda_1}{\alpha})\frac{\gamma^m}{\gamma^m+\mu}
+\tau\beta(1+\frac{\lambda_1}{\alpha})\overset{N}{\underset{j=1}{\sum}}a_{ij}V^*_j}\\
&<\tau\frac{\frac{\beta^m\gamma^m}{\gamma^m+\mu}+\beta\overset{N}{\underset{j=1}{\sum}}a_{ij}V^*_j}
{\lambda_1+\lambda_2+\beta^m(1+\frac{\lambda_1}{\alpha})\frac{\gamma^m}{\gamma^m+\mu}
+\beta(1+\frac{\lambda_1}{\alpha})\overset{N}{\underset{j=1}{\sum}}a_{ij}V^*_j}\\
&=\tau h_{i_0}(V^*)=\tau V^*_{i_0},
\end{align*}
which contradicts the assumption that $U^*_{i_0}=\tau V^*_{i_0}$. Hence, the fixed point is unique. This completes proof.        $\hfill{} \Box$

\section*{Appendix B: Computation of the equilibrium}
In this Appendix, we give the computation process of the equilibrium for the fully-connected network with $N$ nodes.
Assume that $I_i^*=I^{*},i=1,\cdots,N$, then
$$I^{*}=\frac{\frac{\beta^m\gamma^m}{\gamma^m+\mu}+\beta\overset{N}{\underset{j=1}{\sum}}a_{ij}I^{*}}
{\lambda_1+\lambda_2+\beta^m({1+\frac{\lambda_1}{\alpha})\frac{\gamma^m}{\gamma^m+\mu}
}+\beta(1+\frac{\lambda_1}{\alpha})\overset{N}{\underset{j=1}{\sum}}a_{ij}I^{*}}.
$$
Denote $a=\beta(1+\frac{\lambda_1}{\alpha})(N-1)$, $b=(\lambda_1+\lambda_2+m(1+\frac{\lambda_1}{\alpha})-\beta(N-1))$,
$c=-m$, and $m=\frac{\beta^m\gamma^m}{\gamma^m+\mu}$. Then, the above equality can be rewritten as
\begin{equation}\label{forroot}
  a{I^*}^2+bI^*+c=0.
\end{equation}

As $\frac{c}{a}<0$, it follows from the Hurwitz criterion \cite{criterion} that Eq. (\ref{forroot}) has two opposite sign roots.
It is easy  to verify that the positive root $I^*_+$ satisfies $0<I^*_+<1$, so $I^*_+$ is the equilibrium due to the uniqueness of solutions.
Since $[\beta(N-1)+m(1+\lambda_1/\alpha)-(\lambda_1+\lambda_2)]^2<b^2-4ac<[\beta(N-1)+m(1+\lambda_1/\alpha)+(\lambda_1+\lambda_2)]^2$,
it follows that
$$
\frac{1}{1+\frac{\lambda_1}{\alpha}}-\frac{\lambda_1+\lambda_2}{\beta(1+\frac{\lambda_1}{\alpha})(N-1)}
<I^*_+=\frac{-b+\sqrt{b^2-4ac}}{2a}<\frac{1}{1+\frac{\lambda_1}{\alpha}}-\frac{\lambda_1+\lambda_2}{2\beta(1+\frac{\lambda_1}{\alpha})(N-1)}.
$$
Therefore, the equilibrium $I^*\approx \frac{1}{1+\lambda_1/\alpha}$ for the large size fully-connected network.

\begin{figure}[!ht]
\centering
\includegraphics[scale=0.6]{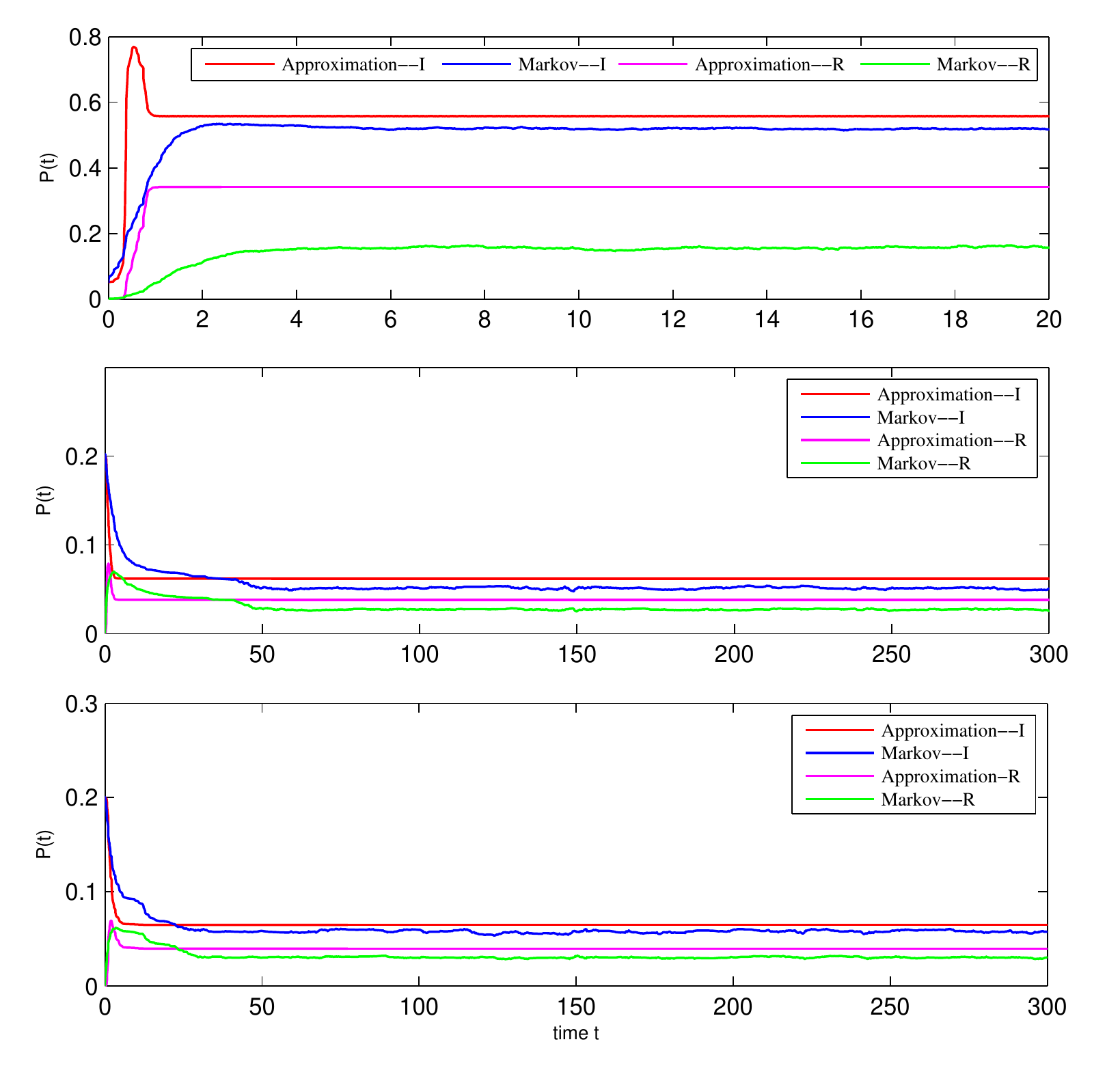}
\caption{\small Comparison of results between two models for fully-connected networks (Top), NW small-world networks (Middle)
and BA scale-free networks (Bottom).}
\label{fig:Performance}
\end{figure}

\section*{Appendix C: Comparison with exact Markov models}
For the purpose of showing the performance of our model,
the following exact Markov model is established by means of continuous-time Markov chain technique \cite{Markov}.
\begin{equation}\label{model0}
\setlength{\abovedisplayskip}{2pt}
\setlength{\belowdisplayskip}{10pt}
\left\{\begin{array}{l}
\frac{dS_i(t)}{dt}=-\beta S_i(t)\overset{N}{\underset{j =1}{\sum}}a_{ij}P\{X_i(t)=0,X_j(t)=1\}-\beta^mS_i(t)I^m(t)+\alpha R_i(t)+\lambda_2I_i(t),\\
\frac{dI_i(t)}{dt}=\beta^mS_i(t)I^m(t)+\beta S_i(t)\overset{N}{\underset{j =1}{\sum}}a_{ij}P\{X_i(t)=0,X_j(t)=1\}-(\lambda_1+\lambda_2)I_i(t),\\
\frac{dR_i(t)}{dt}=\lambda_1I_i(t)-\alpha R_i(t),\\
\frac{dS^m(t)}{dt}=\mu-\mu S^m(t)-\gamma^m S^m(t),\\
\frac{dI^m(t)}{dt}=\gamma^mS^m(t)-\mu I^m(t),
\end{array}\right.
\end{equation}
where $P\{X_i(t)=0,X_j(t)=1\}$ represents the probability of node $i$ being state $R$ and node $j$ being state $I$.

As matter of fact, model (\ref{model0}) turns into model (\ref{model1}) when $P\{X_i(t)=0,X_j(t)=1\}$ will replaced with $I_j(t)$,
namely, the transition rate from state $R$ to state $I$ is linear and equals to $\beta^mI^m(t)+\sum_{i=1}^N a_{ij}I_j(t)$, as shown in Fig.\ref{model}.
The relation between exact Markov model and approximation model please refers to the reference \cite{Abivirus}.

Next, we select three typical networks with 100 nodes,
and use the Gillespite algorithm \cite{exact} to simulate the solution of the Markov model (\ref{model0})
where model parameters and initial conditions are the same as those in Sec. 5.1.
In the experiment, we select randomly initial nodes including susceptible, infected and recovered nodes,
and make 200 realizations for fully-connected networks, and 2000 realizations for NW small-world and BA scale-free networks.

Figure \ref{fig:Performance} shows the comparison of network average states $\bar{I}(t)=\frac{1}{N}\sum_{i=1}^N I_i(t)$ and
$\bar{S}(t)=\frac{1}{N}\sum_{i=1}^N S_i(t)$ between the Markov model and the approximation model.
As time goes, both states $\bar{I}(t)$ and $\bar{S}(t)$ of approximation models are able to describe those of Markov models,
although there is a little overestimation. Furthermore, the estimation for small-world and scale-free networks is better than that for the fully connected networks. On the whole, the performance of this new model is good for describing the real Markov model.

%


\section*{Appendix D: Proof of Theorem 3}
Proof of Theorem 3: Denote
$$\phi_i(\bm{I}^*, \beta^m, \alpha, \beta, \lambda_1)=[\lambda_1+\lambda_2+(1+\frac{\lambda_1}{\alpha})(m+
\beta\overset{N}{\underset{j =1}{\sum}}a_{ij}I_j^*)]I_i^*-m-\beta\overset{N}{\underset{j =1}{\sum}}a_{ij}I_j^*, i = 1, 2, \cdots,N$$
where $m = \frac{\beta^m \lambda^m}{\gamma^m+\mu},\ \bm{I}^*=(I_1^*,...,I_N^*)^T$. Then, it follows from the formula of equilibria in model (\ref{model2}) that
\begin{equation}\label{Eq-Phi}
\phi_i(\bm{I}^*, \beta^m, \alpha, \beta, \lambda_1)=0, \quad i=1,2,\cdots,N
\end{equation}

Taking the partial derivatives of $\phi_i$ with respect to $I_j^*$ and $\alpha$, respectively, one gets
$$J\triangleq(\frac{\partial \phi_i}{\partial (I_j^*)})_{N\times N}=diag(\frac{\lambda_1+\lambda_2}{1-I_i^*(1+\frac{\lambda_1}{\alpha})})+\beta diag((1+\frac{\lambda_1}{\alpha})I_i^* - 1)A,$$
$$\frac{\partial \phi}{\partial \alpha}= -\frac{1}{\alpha^2}*[m\lambda_1E_N+diag(\beta\lambda_1 I_i^*)A]\bm{I}^*.$$
Here $diag(\bullet_i)=diag\{\bullet_1,\bullet_2,\cdots,\bullet_N\}$,
$\frac{\partial \phi}{\partial (\bullet)}=(\frac{\partial \phi_1}{\partial (\bullet)},\cdots,\frac{\partial \phi_N}{\partial (\bullet)})^T$,
$\frac{\partial \bm{I}^*}{\partial (\bullet)}=(\frac{\partial I_1^*}{\partial (\bullet)},\cdots,\frac{\partial I_N^*}{\partial (\bullet)})^T$
and the same below.

According to the implicit differentiation formula, it follows that
$$J\frac{\partial \bm{I}^*}{\partial \alpha}=-\frac{\partial \phi}{\partial \alpha}.$$
Obviously, $\frac{\partial \phi}{\partial \alpha}<0$, namely, $\frac{\partial \phi_i}{\partial \alpha}<0\,(i=1,2,\cdots,N)$.
Next, we prove that $J$ is invertible and all the elements of $(-J)^{-1}$ are negative.

It is easy to obtain that $(1+\frac{\lambda_1}{\alpha})I_i^*-1<0\,(i=1,...,N)$ due to
$$\phi_i=(\lambda_1+\lambda_2)I_i^* + [(1+\frac{\lambda_1}{\alpha})I_i^* - 1](m+
\beta\overset{N}{\underset{j =1}{\sum}}a_{ij}I_j^*)=0.$$ Denote
$M = \beta A - diag(\frac{\lambda_1+\lambda_2}{1-(1+\frac{\lambda_1}{\alpha})I_i^*}) + diag(\frac{m}{I_i^*})$,
and $M_1=M + \max_i\{\frac{\lambda_1+\lambda_2}{1-(1+\frac{\lambda_1}{\alpha})I_i^*}\}E_N$.
Obviously, $M$ is non-negative matrix, and it is irreducible due to the fact that $A$ is irreducible on account of the connectedness of the graph $G$.

According to the Perron-Frobenius Theorem \cite{Perron}, $M_1$ has a simple positive eigenvalue $\rho(M_1)$ and a positive eigenvector $\bm{u}$,
such that
$M_1\bm{u}=\rho(M_1)\bm{u}$. So $M \bm{u}=[\rho(M_1)-\max_i\{\frac{\lambda_1+\lambda_2}{1-(1+\frac{\lambda_1}{\alpha})I_i^*}\}]\bm{u}$,
implying that $\bm{u}$ is also eigenvector of $M$. On the other hand, it follows from Eq.(\ref{Eq-Phi}) that $M\bm{I}^*=0$, indicating that $\bm{I}^*$ is a positive eigenvector of $M$ belonging to eigenvalue $0$. As $\bm{u}^T \bm{I}^* >0\,(\neq 0)$, combining the simplicity of $\rho(M)$ resulted from the simplicity of $\rho(M_1)$, one gets $\rho(M)=0$, indicating that $M$ has a zero eigenvalue and the other eigenvalues are negative.

On the other hand, $-J=M-\beta diag((1+\frac{\lambda_1}{\alpha})I_i^*)A-diag(\frac{m}{I_i^*})$. It follows that all the eigenvalues of matrix $-J$ are negative, and $-J$ is Metzler and irreducible. According to the result in \cite{impact},
all the elements of matrix $(-J)^{-1}$ are negative. Thus
$$\frac{\partial \bm{I}^*}{\partial \alpha}=(-J)^{-1}\frac{\partial \phi}{\partial \alpha}>0,$$
and consequently $\frac{\partial \bar{I}^*}{\partial \alpha}=\frac{1}{N}\sum_{i=1}^N\frac{\partial I_i^*}{\partial \alpha}>0$.

Similarly, we get
$$\frac{\partial \bm{I}^*}{\partial \beta}=(-J)^{-1}\frac{\partial \phi}{\partial \beta}>0,
\frac{\partial \bm{I}^*}{\partial \beta^m}=(-J)^{-1}\frac{\partial \phi}{\partial \beta^m}>0,\;
\mbox{and}\;
\frac{\partial \bm{I}^*}{\partial \lambda_1}=(-J)^{-1}\frac{\partial \phi}{\partial \lambda_1}<0
$$
where $\frac{\partial \phi}{\partial \beta}=diag((1+\frac{\lambda_1}{\alpha})I_i^*-1)A\bm{I}^*$,
$\frac{\partial \phi_i}{\partial \beta^m}=\frac{\lambda^m}{\gamma^m+\mu}[(1+\frac{\lambda_1}{\alpha})I_i^*-1]$, and
$\frac{\partial \phi}{\partial \lambda_1}=[(1+\frac{m}{\alpha})E_N + \frac{\beta}{\alpha}diag(I_i^*)A]\bm{I}^*$.
Therefore, $\frac{\partial \bar{I}^*}{\partial \beta}>0$, $\frac{\partial \bar{I}^*}{\partial \beta^m}>0$,
and $\frac{\partial \bar{I}^*}{\partial \lambda_1}<0$. The proof of Theorem 3 is completed.


\begin{thebibliography}{99}

\bibitem{Pastor2001}Pastor-Satorras R, Vespignani A. Epidemic spreading in scale-free networks[J]. Physical review letters, 2001, 86(14): 3200-3203.

\bibitem{Mishra2007}Mishra B K, Saini D K. SEIRS epidemic model with delay for transmission of malicious objects in computer network[J]. Applied Mathematics and Computation, 2007, 188(2): 1476-1482.

\bibitem{Yuan2008}Yuan H, Chen G. Network virus-epidemic model with the point-to-group information propagation[J]. Applied Mathematics and Computation, 2008, 206(1): 357-367.

\bibitem{Newman2002}Newman M E J. Spread of epidemic disease on networks[J]. Physical review E, 2002, 66(1): 016128.


\bibitem{Selley2015}S\'{e}lley F, Besenyei \'{A}, Kiss I Z, et al. Dynamic control of modern network-based epidemic models[J]. SIAM Journal on applied dynamical systems, 2015, 14(1): 168-187.

\bibitem{Liu2017}Liu Q, Sun M, Li T. Analysis of an SIRS epidemic model with time delay on heterogeneous network [J]. Advances in Difference Equations, 2017, 2017(1): 309.



%



\bibitem{Wei2015}Wei X, Liu L, Zhou W. Global stability and attractivity of a network-based SIS epidemic model with nonmonotone incidence rate[J]. Physica A: Statistical Mechanics and its Applications, 2017, 469: 789-798.



\bibitem{WeiWu2018}Xiang Wei, Xiaoqun Wu, Shihua Chen, Jun-an Lu, Guanrong Chen, Cooperative epidemic spreading on a two-layered interconnected network, SIAM Journal Applied Dynamical Systems, 2018, 17(2): 1503-1520.

\bibitem{Pastor2015}Pastor-Satorras R, Castellano C,Mieghem P, Vespignani A. Epidemic processes in complex networks. Reviews of Modern Physics 2015, 87(3):925-979.

\bibitem{Mieghem2009}Mieghem PV, Omic J, Kooij R. Virus spread in networks. IEEE/ACM Transactions on Networking 2009, 17(1):1-14.

\bibitem{YoussefM2011}Youssef M, Scoglio C. An individual-based approach to SIR epidemics in contact networks[J]. Journal of theoretical biology, 2011, 283(1): 136-144.

\bibitem{Plos one2015}Yang L X, Draief M, Yang X. The impact of the network topology on the viral prevalence: a node-based approach[J]. PloS one, 2015, 10(7): 0134507.

\bibitem{LX Yang2017}Yang L, Draief M, Yang X. Heterogeneous virus propagation in networks: a theoretical study[J]. Mathematical Methods in the Applied Sciences, 2017, 40(5): 1396-1413.

\bibitem{Shi2008}Shi H, Duan Z, Chen G. An SIS model with infective medium on complex networks[J]. Physica A: Statistical Mechanics and its Applications, 2008, 387(8-9): 2133-2144.

\bibitem{Meng Yang2011}Yang M, Chen G, Fu X. A modified SIS model with an infective medium on complex networks and its global stability[J]. Physica A: Statistical Mechanics and its Applications, 2011, 390(12): 2408-2413.

\bibitem{Wang2012}Wang Y, Jin Z, Yang Z, et al. Global analysis of an SIS model with an infective vector on complex networks[J]. Nonlinear Analysis: Real World Applications, 2012, 13(2): 543-557.

\bibitem{Define}Hom R A, Johnson C R. Topics in matrix analysis[M]. Cambridge UP, New York, 1991.

\bibitem{HurwitzMet}Narendra K S, Shorten R. Hurwitz stability of Metzler matrices[J]. IEEE Transactions on Automatic Control, 2010, 55(6): 1484-1487.

\bibitem{compact}Yorke J A. Invariance for ordinary differential equations[J]. Mathematical systems theory, 1967, 1(4): 353-372.

\bibitem{Brower}Shamash E R. Fixed point theory: Banach, Brouwer and Schauder theorems[D]. California State University, Northridge, 2000.

\bibitem{criterion}Robinson R C. An introduction to dynamical systems: continuous and discrete[M]. American Mathematical Soc., 2012.

\bibitem{yangan2005}Yan G, Zhou T, Wang J, Fu Z Q, Wang B H. Epidemic spread in weighted scale-free networks. Chinese Phys. Lett. 2005, 22:510-513.

\bibitem{Abivirus}Yang L X, Yang X, Tang Y Y. A bi-virus competing spreading model with generic infection rates[J]. IEEE Transactions on Network Science and Engineering, 2018, 5(1):2-13.

\bibitem{exact}Gillespie D T. Exact stochastic simulation of coupled chemical reactions[J]. The journal of physical chemistry, 1977, 81(25): 2340-2361.
\bibitem{impact}Yang L X, Yang X, Wu Y. The impact of patch forwarding on the prevalence of computer virus: a theoretical assessment approach[J]. Applied Mathematical Modelling, 2017, 43: 110-125.

\bibitem{Markov}Stewart W J , Probability, Markov Chains, Queues, and Simulation: The Mathematical Basis of Performance Mdeling [M], Princeton University Press, 2009.

\bibitem{Perron}Bhatia R, Matrix Analysis[M], Springer-Verlag, New York, USA, 2011 .










\end{thebibliography}
\end{document}